# Investigating Decision Support Techniques for Automating Cloud Service Selection

Miranda Zhang[1,2]

Supervised by: Rajiv Ranjan[1], Armin Haller[1], Dimitrios Georgakopoulos[1], Peter Strazdins[2]

[1] Information Engineering Laboratory, CSIRO ICT Centre
{miranda.zhang, rajiv.ranjan, armin.haller , dimitrios.georgakopoulos }@csiro.au
[2] Research School of Computer Science, ANU
Peter.Strazdins@cs.anu.edu.au, miranda.zhang@anu.edu.au

*Abstract*—**The compass of Cloud infrastructure services advances steadily leaving users in the agony of choice. To be able to select the best mix of service offering from an abundance of possibilities, users must consider complex dependencies and heterogeneous sets of criteria. Therefore, we present a PhD thesis proposal on investigating an intelligent decision support system for selecting Cloud-based infrastructure services (e.g. storage, network, CPU). The outcomes of this will be decision support tools and techniques, which will automate and map users' specified application requirements to Cloud service configurations.**

## I. MOTIVATION

The emergence of Cloud computing [1] [2] [3][29] over the past five years is potentially one of the breakthrough advances in the history of computing. The Cloud computing paradigm is shifting computing from in-house managed hardware and software resources to virtualized Cloud-hosted services. Cloud computing assembles large networks of virtualized services: hardware resources (CPU, storage, and network) and software resources (e.g., web server, databases, message queuing systems, monitoring systems.). Cloud service types can be abstracted into three layers: Software as a Service (SaaS), Platform as a Service (PaaS), and Infrastructure as a Service (IaaS). Hardware and software resources form the basis for delivering IaaS and PaaS. The top layer focuses on application services (SaaS) by making use of services provided by the lower layers. PaaS/SaaS services are often developed and provided by third party service providers who are different from the IaaS providers.

Cloud providers including Amazon Web Services (AWS), Microsoft Azure, Rackspace, GoGrid, and others give users the option to deploy their application over a pool of virtually infinite services with practically no capital investment and with modest operating costs proportional to the actual use. Elasticity, cost benefits and abundance of resources motivate many organizations to migrate their enterprise applications to the Cloud. Although Cloud offers the opportunity to focus on revenue growth and innovation, decision makers (e.g., CIOs, scientists, developers, engineers, etc.) are faced with the complexity of choosing the right service delivery model for composite application and infrastructure across private, public, and hybrid Clouds.

With Cloud providers and service offerings having grown in numbers, the migration of applications (e.g., multi-layered enterprise application, scientific experiments, video-on-demand streaming application, IPTV, etc.) to the Cloud demands selecting the best mix of services across multiple layers (e.g., IaaS, PaaS, and SaaS) from an abundance of possibilities. Any such Cloud service selection decision has to cater for a number of conflicting criteria, e.g. throughput and cost, while ensuring that Quality of Service (QoS) requirements are met. The problem is further aggravated by the fact that different applications have heterogeneous QoS requirements. For example, QoS requirements for scientific experiments (e.g., deadline) differ from video-on-demand streaming application (e.g., streaming latency, resolution, etc.).

Existing service selection methods have not kept pace with the rapid emergence of the multiple-layer nature of Cloud Services. Notably, techniques for web service selection and grid job scheduling [28] cannot be adopted for Cloud Service Selection, because they do not cater for the diverse sets of criteria and their dependencies across multiple layers of Cloud Services. Further, current selection approaches are rarely transparent and adaptive. They require the user to have familiarity with the various Cloud Providers and extensive hard-coded and static scripts. This is inadequate, given the proliferation of new providers offering services at different layers.

## II. RESEARCH PROBLEMS

We have identified the following hard research problems in the domain of Cloud Service Selection and Comparison (CSSC).

### Q1. Automatic service identification and representation

A cumbersome task for decision makers is to manually read Cloud providers' documentation for finding out which services are suitable for building their Cloud-based application architecture (e.g., a biologist intending to host his genomics experiment in the Cloud). This problem is further aggravated due to the rapid emergence of services in the Cloud landscape. The multi-layered organization (e.g., SaaS, PaaS, and IaaS) of Cloud Services, along with their heterogeneous types (CPU, Storage, Network, web server, databases, etc.) and features (Virtualization technology, SLA model, billing model, Cloud location, cost, etc.) makes the task of service identification a hard problem. In addition, the use of non-standardized naming terminology used by Cloud providers makes this problem challenging. For example, AWS refers to CPU services as Elastic Compute Cloud (EC2) Unit while GoGrid refers to the same as Cloud Servers. Cloud providers typically publish their service description on their websites in various layouts they prefer. The relevant information may be updated without prior notice to the users. Further, the structure of web pages can change significantly leading to confusion. Hence, it is not an easy task to obtain reliable service descriptions from Cloud providers' website and documentation (which are the only sources of information). This leads to the following challenges: How to automatically fetch service description published by Cloud providers and present them to decision makers in a human readable way? Can we develop a unified and generic Cloud ontology to describe the services of any Cloud provider which exists now or may become available in the future?

*Q2. Optimized Cloud Service Selection and Comparison*

Matching results to decision makers' requirements involves bundling of multiple related Cloud services, computing combined cost (under different billing models and discount offers), considering all possible (or only valuable) alternatives and multiple selection criteria (including specific features, long-term management issues and versioning support). The diversity of offerings in the Cloud landscape leads to practical research questions: how well does a service of a Cloud provider perform compared to the other providers [3] [4] [5]? Which Cloud services are compatible to be combined or bundled together [1] [10][11]? How to optimize the process of composite Cloud service selection and bundling? For example, how does a decision maker compare the cost/performance features of CPU, storage, and network service offered by AWS EC2, Microsoft Azure, GoGrid, FelxiScale, TerreMark, and RackSpace. Though branded calculators are available from individual Cloud providers for calculating service leasing cost, it is not easy for decision makers to generalise their requirements to fit different service offers (with various quota and limitations) let alone computing and comparing costs. For instance, a low-end CPU service of Microsoft Azure is 30% more expensive than the comparable AWS EC2 CPU service, but it can process an application workload twice as quickly. Similarly, a decision maker may choose one provider for storage intensive applications and another for computation intensive applications.

*Q3. Simplified interfaces for Cloud Service Selection*

Despite the popularity of Cloud Computing, existing Cloud Service manipulations (e.g. select, start, stop, configure, delete, scale and de-scale) techniques require human familiarity with different Cloud service types and typically rely on procedural programming or scripting languages. The interaction with services is performed through low-level application programming interfaces (APIs) and command line interfaces. This is inadequate, given the proliferation of new providers offering services at different layers (e.g. SaaS, PaaS, and IaaS). One of the consequences of this state is that accessibility to Cloud Computing is limited to decision makers with IT expertise. This raises a set of research questions: How to develop interfaces that can transform low, system-level programming to easy-to-use drag and drop operations? Will such interfaces improve and simplify the process of CSSC?

## III. PROPOSED APPROACH AND METHODOLOGY

The research required for Q1 has been partially addressed and is discussed in section VI. Our approach to solve Q2 involves the investigation of key criteria and their semantics to explore and select Cloud services. While there is a growing interest in this area, the set of concepts needed to understand the selection problem is still emerging (discussed in section V).We will use detailed case studies by surveying decision makers who are considering – or may consider – the migration of applications to Clouds. We will clearly identify the most important selection criteria and Cloud Service alternatives, considering different application use-cases. The choice of a particular selection methodology stems from the semantics (quantitative vs. qualitative) of criteria and the number of feasible alternatives (Cloud Services) and their types (IaaS, PaaS, or SaaS). Given the nature of the selection problem, it can be safely concluded that optimization techniques such as linear and non-linear programming, may not be effective, as they are incapable of handling multiple conflicting criteria (e.g., maximize throughput and utilization; minimize cost and latency) and they have known scalability issues when there are many alternatives, as found in large, heterogeneous mixes of services in Clouds.

To solve Q2, we will propose and develop a novel and flexible decision-making framework that builds upon two distinct techniques: i) evolutionary optimization techniques, the process of simultaneously optimizing two or more conflicting objectives expressed in the form of linear or non-linear functions of criteria; ii) a decision making method, attempting to identify and select alternatives based on the value and the goals of decision makers. Genetic Algorithms (GA) such as the Non-dominated Sorting Genetic Algorithm-II and Strength Pareto Evolutionary Approach-2 are the most commonly used multiple criteria optimization techniques. Since these are, by their nature, unconstrained procedures, it is necessary to find a way to integrate penalty functions with fitness functions during the optimization process. On the other hand, they have the capability of handling search over an infinite number of feasible alternatives constrained by a finite number of quantitative criteria. In recognition of the complexity of these techniques, we analyse their computational tractability thoroughly, and derive several time and space complexity bounds to measure the computational quality.

The main limitation of multi-criteria optimization techniques is that they cannot handle mixed qualitative (e.g. hosting region, operating system type) and quantitative criteria. Multi-criteria decision-making techniques, including Analytic Hierarchy Process (AHP) and others can handle mixed qualitative and quantitative criteria. AHP is based on pair-wise comparisons of the criteria. For each pair of criteria, the administrator is required to provide a subjective opinion of their relative importance. Moreover, AHP can only be applied to a finite set of alternatives. Nonetheless, optimization techniques such as GA can use AHP to evaluate alternatives and find an optimal solution according to the

AHP-based evaluation. For example, by combining a GA with AHP a hybrid multi-goal optimization heuristic method can be created. Therefore, AHP can be employed as a fitness function that evaluates individuals regarding multiple criteria. This, however, introduces a few requirements for the structure of individuals, in particular for the genetic representation of the individuals. A new challenge resulting from the combination of GA and AHP is to transfer subjective opinions stated once into an AHP-based fitness function, so a novel approach is required. AHP requires subjective opinions to be stated for each alternative. This becomes unsolvable with a potentially infinite number of alternatives considered in a GA. No results have yet been published on combining optimization methods with decision making methods to enable optimized and flexible selection of Cloud services.

We will evaluate the correctness and effectiveness of our approach by measuring the performance of our prototype system, whether it is able to find the global optimal(s) and all pareto-optimal solutions [12] and its time and space complexity. Note that QoS aware service selection problem [13] is NP-hard which takes exponential time and costs to solve. Hence simplification approaches (e.g. local maximization [15], absence of QoS constrains [16], single objective optimization/simple additive weighting (SAW) [12], space pruning) would be used. We will also compare the trade offs and benefits of each of those.

To solve Q3, we will investigate a widget-based visual programming language to simplify the interaction with Cloud Services. The widget interfaces will not overwhelm the decision maker with excessive input fields (e.g. different configuration features of Cloud Services. However, this will require the decision support service to cope with minimum and invalid inputs from human and computer generated requests. Numerical values do not always make sense and may require too much effort to compute, so vague inputs in addition to exact values should be allowed, and results should be carefully presented in a way that is easy to understand (for example, illustrated with a graph). Decision makers might be interested in aspects which are hard to evaluate, like how good is the customer and technical (API, language) support. Decision makers' rating or reputation tracking techniques can be used in such situations. We will also investigate how historical selection data can be used to derive the popularity of each service (e.g. by tracking how many times each service come up as the recommended solution and/or being selected by users), which can be used for future recommendations.

We will conduct a user satisfaction study by collecting feedback from users with different level of background knowledge, i.e. elementary, medium, expert. Some example questions can be: how easy is it to navigate through the system (number of clicks before getting results), does the result include everything they want to know, or is there too much text/information displayed. We would also calculate the average time for a user to make a Cloud service selection decision with the aid of our system and compare it to the scenario where the user manually browses through the Cloud provider webpage and makes the decision.

## IV. EXPECTED CONTRIBUTIONS AND IMPACT

This PhD research has immediate practical significance and relevance. The International Data Corporation (IDC) forecasts that spending on public IT Cloud-hosted applications (excluding private Cloud hosting) will grow from $16.5 billion in 2009 – a modest, recession-influenced forecast – to over $55 billion in 2014" (27% p.a. growth) [20]. This research addresses important fundamental research issues, and develops a novel end-to-end framework. The outcomes of this research will make significant scientific advances in understanding the theoretical and practical problems in selecting and comparing Cloud services.

The first contribution of this PhD thesis was validated by solving Q1 [17] (described in section VI). In summary, the contribution made by solving Q1 includes: (i) A unified and formalized domain model (see Section VI) capable of fully describing infrastructure services in Cloud computing. The model is based and has been successfully validated against the most commonly available infrastructure services including Amazon, Microsoft Azure, GoGrid, etc and (ii) An implementation of a declarative decision support system which we call as *CloudRecommender* for the selection of infrastructure Cloud service configurations using transactional SQL semantics, procedures and views. The benefits to users of CloudRecommender include, for example, the ability to estimate costs, compute cost savings across multiple providers with possible tradeoffs and aid in the selection of Cloud services [17].

Next, the contributions that we will make by solving Q2 include: (i) formulating computationally tractable fitness and penalty functions pertaining to the selection of Cloud services given the user application use-case (e.g., scientists searching for Cloud services for conducting data-intensive eResearch experiments and developers searching for Cloud services for hosting web applications). Based on the above contribution, we will also implement (ii) a new decision making framework utilizing hybrid optimization techniques to transform primarily (but not limited to) Cloud Service selection from manual and time-consuming scripting to a process that is flexible, and to a large extend automated. This framework (improved version of CloudRecommender – see Section VI) will help decision makers choose the Cloud Service Provider(s) that best fits performance, feature, and cost needs. On the other hand, Cloud providers can apply these techniques to identify their areas of competitive disadvantage and thus lead to redesign and improvement.

The possible contribution by solving Q3 will be simplified user-interfaces, which will allow users from non-IT domain (scientists, business analyst) to select and compare Cloud services. The research therefore contributes to the eResearch domain, which is the application of ICT capabilities to improve efficiencies of fundamental research.

## V. PROGRESS BEYOND STATE OF THE ART

In relation to Q1, there are 3 common approaches for web services identification/publication: 1) manually maintain directories by categorizing manually-submitted or collected information about Cloud services and providers, an example of such kind is Universal Description, Discovery and Integration (UDDI), which has failed to gain wide adoption; 2) use web crawling, and automatically create listings; and 3) combine both, e.g. using manually-submitted URIs as seeds to generate indexes. The first approach is the only feasible solution at the moment. But extensive research and standardization efforts have been put into developing web information representation models, namely, the Resource Description Framework (RDF), the semantic web, and ontologies [19]. Some of the recent research such as [2] has focused on Cloud storage service (IaaS level) representation using XML. But the proposed schema does not comply with or take into account any of the above mentioned standards. We believe that semantic web technologies should be adopted to standardize the Cloud services representations. For example, there is a popular demand for the schema of software services in the Schema.org forum, proposing an extension that specifically defines Cloud services (which includes SaaS) would greatly facilitate service publication and identification using semantic web technologies.

For Q2, a number of research [9] and commercial projects (mostly in their early stages) provide simple cost calculation or benchmarking and status monitoring, but none is capable to consolidate all aspects and provide a comprehensive ranking of infrastructure services. For instance, CloudHarmony provides up-to-date benchmark results without considering cost, Cloudorado calculates the price of IaaS-level CPU services based on static features (e.g., processor type, processor speed, I/O capacity, etc.) while ignoring dynamic QoS features (e.g., latency, throughput, load, utilization, etc.). Yuruware Compare [14] beta version offers elementary search on Compute IaaS Cloud services. Although they aim to provide an integrated tool with monitoring and deploying capabilities, it is still under development and the finish date is unknown. The current version does not allow selection of storage service by itself and QoS has not been compared. Though branded calculators are available from individual Cloud providers, such as Amazon [4], Azure [5], and GoGrid, for calculating service leasing cost, it is not easy for users to generalize their requirements to fit different service offers (with various quota and limitations) let alone computing and comparing costs.

## VI. Preliminary Results

The result of Q1 is summarized in a first conference paper [17] which is under review at the GECON 2012 conference. In this paper, we presented a declarative approach to Cloud service selection, comparison and its implementation as CloudRecommender system. We formally capture the domain knowledge (e.g., IaaS configurations) using a declarative logic-based language, and then apply the knowledge on top of relational data model that encapsulates Cloud-wide information. Based on this knowledge, we have drawn the relationships in the conceptual IaaS configuration model and represented in Fig. 1 and we illustrated the system architecture in Fig. 2. We have applied declarative service selection technique by utilizing SQL and regular expressions minimize side effects and reinforce constrains. Thus, leading to a improved Cloud service representation and selection.

The service selection logic developed by our research is transactional and applies well-defined SQL semantics for querying, inserting, and updating IaaS configurations. In addition, the proposed declarative approach is preferable over hard coding the sorting and selection algorithm (as used in [6]) as it allows us to take the advantage of optimized query operations (e.g. select and join). The problem we are trying to solve involves computing the Cartesian product $O(N * M)$ of multiple sets of options. A widely used solution of such operation is the JOIN operation in database. Note that much work in database-systems has aimed at efficient implementation of joins. In fact, modern database often use HASH JOIN $O(N + M)$ and MERGE JOIN $O(N*Log(N) + M*Log(M))$. They are much faster than $O(N * M)$.

Prior to CloudRecommender, there have been a variety of systems that use declarative logic-based techniques for managing resources in distributed computing systems. The focus of the authors in work [21] is to provide a distributed platform that enables Cloud providers to automate the process of service orchestration via the use of declarative policy languages. The authors in [7] present an SQL-based decision query language for providing a high-level abstraction for expressing decision guidance problems in an intuitive manner so that database programmers can use mathematical programming technique without prior experience. We draw a lot of inspiration from the work in [22] which proposes a data-centric (declarative) framework to improve SLA fulfillment ability of Cloud service providers by dynamically relocating infrastructure services. COOLDAID [8] presents a declarative approach to manage configuration of network devices and adopts a relational data model and Datalog-style query language. NetDB [23] uses a relational database to manage the configurations of network devices. However, NetDB is a data warehouse, not designed for Cloud service selection or comparison. Puppet [18] manages the configuration of data-centre resources using a custom and user-friendly declarative language which simplifies the management of data centre resources for providers.

In contrast to the aforementioned approaches, CloudRecommender is designed for solving the new challenge of handling heterogeneous service configuration and naming conventions in Cloud computing. It is designed with a different application domain – one that aims to apply declarative (SQL) and widget programming technique for solving the Cloud service selection problem.

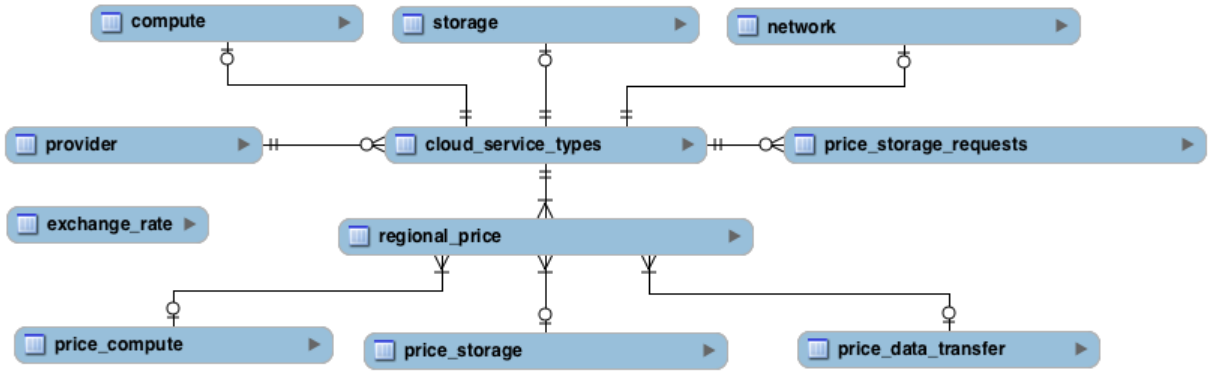

Figure 1. Conceptual UML data model representing infrastructure service entities and their relationships

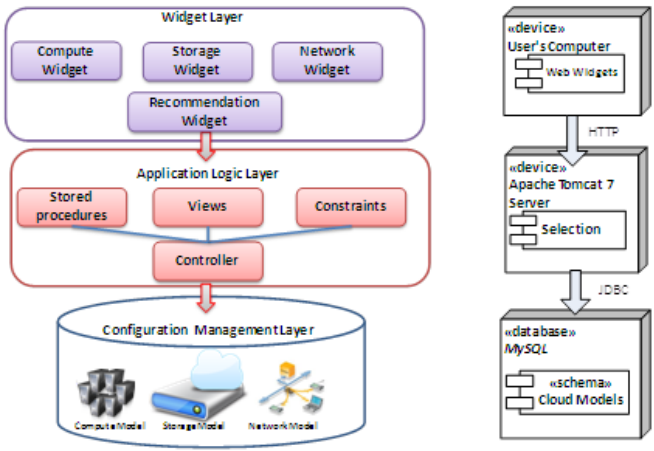

Figure 2. System architecture and deployment structure

CloudRecommender also exposes REpresentational State Transfer (RESTful) APIs that help external applications to programmatically obtain results, i.e. recommended infrastructure Cloud services configurations, shown in Fig. 5.

## VII. MOTIVATIONAL EXAMPLES

Gaia is a global space astrometry mission with a goal of making the largest, most precise three-dimensional map of our Galaxy by surveying more than one billion stars. For the amount of images produced by the satellite (1 billion stars x 80 observations x 10 readouts), if it took one millisecond to process one image, it would take 30 years of data processing time on a single processor. Luckily the data does not need to be processed continuously, every 6 months they need to process all the observations in as short a time as possible (typically two weeks) [24]. Hypothetically speaking say they choose to use 120 high CPU VMs. Example search via CloudRecommender version 1 is shown in Fig. 3. With each VM running 12 threads, there were 1440 processes working in parallel. This will reduce the processing time to less than 200 hours (about a week).

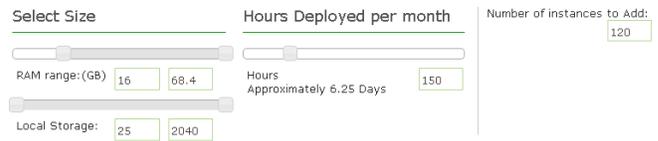

Figure 3. Example input parameter values

In this case since data can be moved into/out of the cloud in bulk periodically, FedEx hard drive may be preferred over transferring data over the internet.

Promotional offers may not matter much in this case compare to the huge time and capital investment savings. But it makes a big difference for small business (or start ups) running a website.

Another example usage is sites with large continuous data input and processing need like Yelp. Everyday Yelp generates and stores around 100GB of logs and photos; runs approximately 200 MapReduce jobs and processing 3TB of data [25]. Yelp.com has more than 71 million monthly unique visitors [26]. The average page size of a typical website is about 784 kB [27]. So the estimated data download traffic is about 51TB per month if every unique user only views one page once a month. Fig. 4 shows a sample search for the above mentioned scenario.

Figure 4. Example parameters for REST API

Figure 5.   REST call via HTTP GET request